\documentclass[conference]{IEEEtran}
\IEEEoverridecommandlockouts
\usepackage{cite}
\usepackage{amsmath,amssymb,amsfonts}
\usepackage{algorithmic}
\usepackage{graphicx}
\usepackage{textcomp}
\usepackage{xcolor}
\usepackage{times}
\usepackage{soul}
\usepackage{url}
\usepackage[hidelinks]{hyperref}
\usepackage[utf8]{inputenc}
\usepackage[small]{caption}
\usepackage{amsthm}
\usepackage[ruled,vlined]{algorithm2e}
\usepackage{booktabs}
\usepackage{array}
\usepackage{placeins}
\def\BibTeX{{\rm B\kern-.05em{\sc i\kern-.025em b}\kern-.08em
    T\kern-.1667em\lower.7ex\hbox{E}\kern-.125emX}}

\begin{document}

\title{Interactive Explanations: Diagnosis and Repair of Reinforcement Learning Based Agent Behaviors}

\author{\IEEEauthorblockN{Christian Arzate Cruz}
\IEEEauthorblockA{\textit{The University of Tokyo} \\
Tokyo, Japan \\
arzate.christian@gmail.com}
\and
\IEEEauthorblockN{Takeo Igarashi}
\IEEEauthorblockA{\textit{The University of Tokyo} \\
Tokyo, Japan \\
takeo@acm.org}
}

\IEEEoverridecommandlockouts
\IEEEpubid{\makebox[\columnwidth]{\copyright20XX IEEE \hfill} \hspace{\columnsep}\makebox[\columnwidth]{ }}

\maketitle

\IEEEpubidadjcol

\begin{abstract}
Reinforcement learning techniques successfully generate convincing agent behaviors, but it is still difficult to tailor the behavior to align with a user's specific preferences. What is missing is a communication method for the system to explain the behavior and for the user to repair it. In this paper, we present a novel interaction method that uses interactive explanations using templates of natural language as a communication method. The main advantage of this interaction method is that it enables a two-way communication channel between users and the agent; the bot can explain its thinking procedure to the users, and the users can communicate their behavior preferences to the bot using the same interactive explanations. In this manner, the thinking procedure of the bot is transparent, and users can provide corrections to the bot that include a suggested action to take, a goal to achieve, and the reasons behind these decisions. We tested our proposed method in a clone of the video game named \textit{Super Mario Bros.}, and the results demonstrate that our interactive explanation approach is effective at diagnosing and repairing bot behaviors.
\end{abstract}


\section{Introduction}
\begin{figure}
    \begin{center}
        \includegraphics[width = 0.9 \columnwidth ]{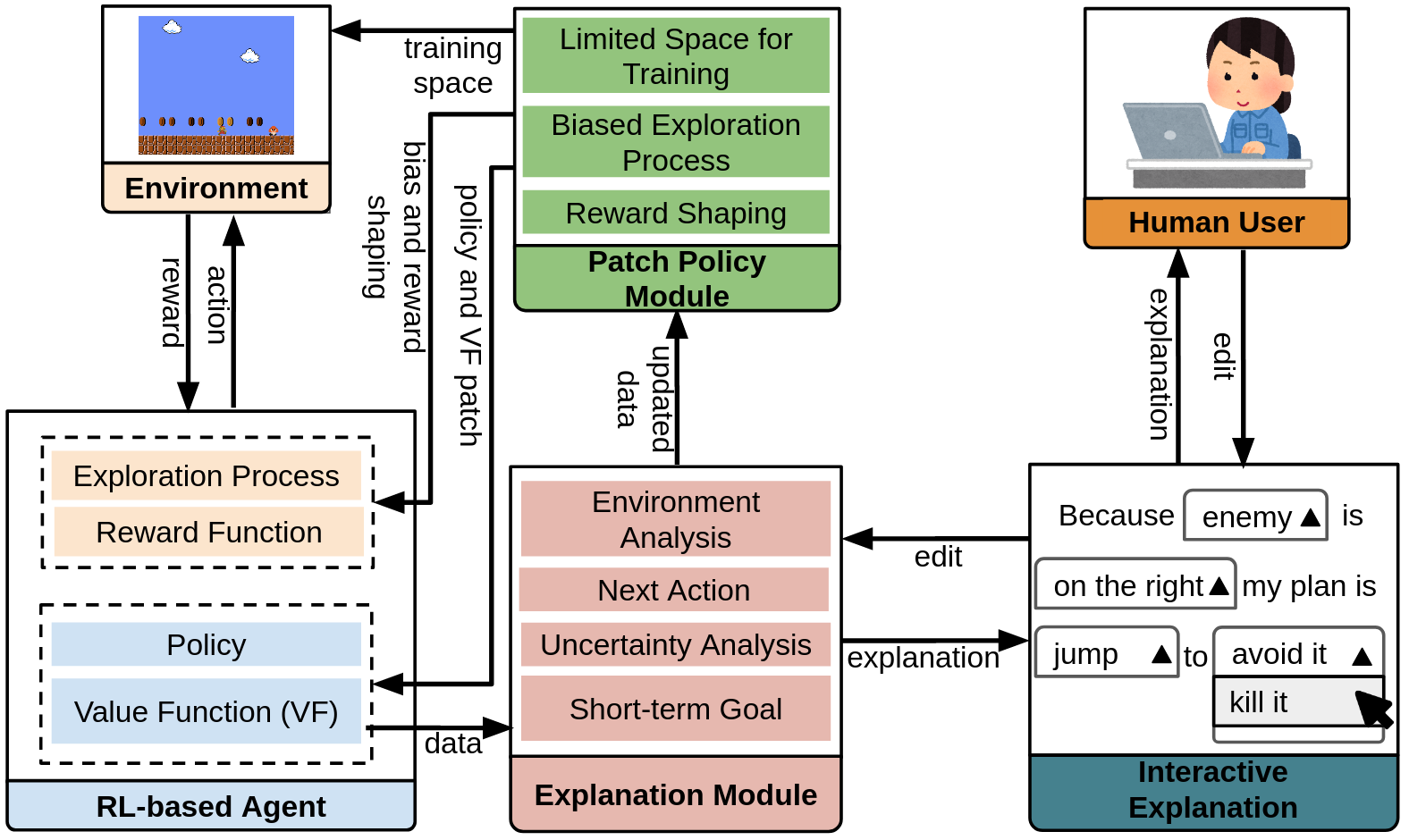}
        \caption{Our interactive explanations framework.}
        \label{fig::interactiveExplanation}
    \end{center}
    \end{figure}

Creating agents using reinforcement learning (RL) techniques is a research area that has seen many advances in recent years but many challenges remain. Arguably, one of the main obstacles in RL research is the agent alignment problem \cite{Leike2018}; this problem arises when we try to create agents that act as the users envision \cite{Ilhan2019,Kanervisto2020}. In this paper, we focus on one of the main difficulties of the agent alignment problem: facilitating the diagnosis and repair of unacceptable outcomes while minimizing the need for feedback from users.

Creating RL-based bots that extract interesting elements regarding their thinking procedure is an effective way to diagnosis the cause of bugs in their policy \cite{Sequeira2019}. Also, in the same direction, we find works that enable RL agents to explain their behavior by contrasting the outcomes of multiple policy options \cite{Waa2018,Hayes2017}.  For this work, we developed a bot that combines both aforementioned approaches; our bot can explain its thinking procedure and can compare the results of multiple policies and present the contrasting outcomes to the user. Furthermore, we extend the scope of the explanations by providing information about the uncertainty of the results after taking a particular action, and what goal the bot is trying to achieve in the next few time-steps. This additional information in the explanations is vital for understanding the accuracy of the transition model of the bot, and to better understand how the reward function affects the policy at a given state. 

To repair the behavior of a bot, we make templates of natural language explanation interactive for the user to give feedback to the bot (see Figure \ref{fig::interactiveExplanation}). With this novel interaction procedure, the user provides corrections that include a suggested action to take, a goal to achieve, and the reasons behind these decisions. In the literature, we can find works that map natural language to a reward function \cite{Williams2018}, for describing goals \cite{Misra2017}, or learning about the dynamic of the environment \cite{Narasimhan2018}. In contrast to these approaches, we used the explanations to design constrained testbeds for our bot to train in with a biased exploration process. After this attention-based exploration process, we compute a value function and policy, which we apply to all the states in the environment that match the description given by the user. Our method provides fast learning cycles that let the users observe the results of their feedback after just a few seconds.

We evaluated our proposed method in a clone of the video game named \textit{Super Mario Bros}. In our user test with $13$ non-experts in RL with varied backgrounds (design, humanities, and computer science), we demonstrated that users were able to diagnose and repair the policy of the bot. Moreover, users were able to adapt the play style of the bot according to their preferences. This empirical evidence suggests that our interactive explanation method is an economical and pragmatic alternative to tackling the agent alignment problem. Besides, since our system is based on the Markov Decision Process (MDP) framework, it would be relatively easy to adapt our method for different applications that could be useful in modern game development such as synthetic testers \cite{Holmgaard2018,Suetake2020}, human-like bots \cite{Broll2019,Pelling2019}, and procedural content generation via RL \cite{Khalifa2020}.

\section{Related Work}
Research on explainable artificial intelligence  \cite{Holzinger2018machine,Xu2019toward,Tao2020} and agents that learn from natural language \cite{li2019pumice,Little2006translating} is extensive. For brevity, in the rest of this section, we focus on contrasting our work against current related research on these subjects that specifically use the reinforcement learning (RL) framework as a basis.

\subsection{Interactive Reinforcement Learning (RL)}
For our work, we follow an interactive RL setting since we have a human-in-the-loop that tailors the underlying RL algorithm to improve, or personalize, the policy of the bot. According to the classification by \cite{Arzate2020}, our work fits the design dimension that focuses on adapting the exploration process of the bot. The most similar implementations to ours are the goal biasing \cite{Suay2011} and action biasing \cite{Yu2018} approaches. 

\cite{Suay2011} propose a method that directs the agent's attention towards an object of interest on the screen; the exploration bias is driven by selecting actions that will get the agent close to the selected object (goal). \cite{Yu2018} present an algorithm that biases the exploration process based on the binary feedback from the user; that is, the agent tends to perform the actions that the user evaluated as good over those marked as bad. 

The main difference between our biased exploration process and the work in \cite{Suay2011} and \cite{Yu2018} is that using all the information from the interactive explanations, we generate a small training environment where the bot learns how to achieve the proposed goal by exploring it and following a strategy that is biased by the suggested action. Once finished the exploration process, we compute a value function and policy that we integrate into the main policy (used in all the state-space) in the states that suit the description given by the user in the interactive explanation. In this manner, we minimize the required feedback from users since we generalize their feedback to all the similar states in the environment.   

Another difference is that \cite{Suay2011} uses the goal to bias the decision-making of the bot to take actions that lead it to the suggested object/goal. Differently, in our method, we use the proposed goal for reward shaping. We carry out the reward shaping method by suppressing all the reward signals except for the one that the user is proposing. Our approach has the advantage of not requiring a precise model of the dynamics of the environment to effectively choose actions that will take to bot closer to the goal. Besides, we use the action as advice that biases the exploration rather than a critique like \cite{Yu2018} do.

For a survey on interactive RL, we refer our reader to \cite{Arzate2020}.

\subsection{Explainable RL}
We build our interactive explanation based on the work by \cite{Sequeira2019}. They propose a three-level introspection procedure for RL agents that extracts interesting elements from the agent's behavior and its interactions with the environment. For the first level, the bot analyzes the transitions and rewards of its underlying Markov Decision Process (MDP). In the second level, the analysis focuses on the history of interactions with the environment. For the third level, a meta-analysis combines elements generated by the previous levels. We implement these three levels of introspection: we provide the users with information about the environment, the task that the bot is trying to solve, the interaction between the bot and the environment, and an analysis of the current goal. Furthermore, we complement our interactive explanation with an interrogative analysis similar to the work by \cite{Ko2004}. This analysis empowers users with the ability to ask ``why'' and ``why not'' questions. We use these questions to form explanations that contrast the result against one particular action (``why not'') or against all possible actions (``Why''). Moreover, in the interactive explanations we include information about how safe (or unsafe) performing a specific action is at a given state. Users can use this uncertainty information to get a better idea of the model of the bot or to correct it by manually changing its value. 

\subsection{Using Natural Language to Aid RL}
When we use our interactive explanations as input, users aid the agent with natural language templates. \cite{Misra2017}  propose an approach, similar to ours, to train RL agents through reward shaping by specifying the goal-states with natural language templates. Similarly, \cite{Kuhlmann2004} map natural language to a set of rules that increase or decrease the probability of selecting specific actions during training in an RL setting. On the other hand, our interactive explanations approach provides users with a natural language template that lets them specify more elements besides goals or preferred action. Additionally, using our interactive explanation to tailor the elements of the underlying RL algorithm allows us to create fixing patches for the main policy in a fast manner, which is vital to have a good user experience.

We refer our readers to \cite{Luketina2019} for a survey on RL informed by natural language.

\section{Interactive Explanations}
\begin{figure}
\begin{center}
    \includegraphics[width = 0.95 \columnwidth ]{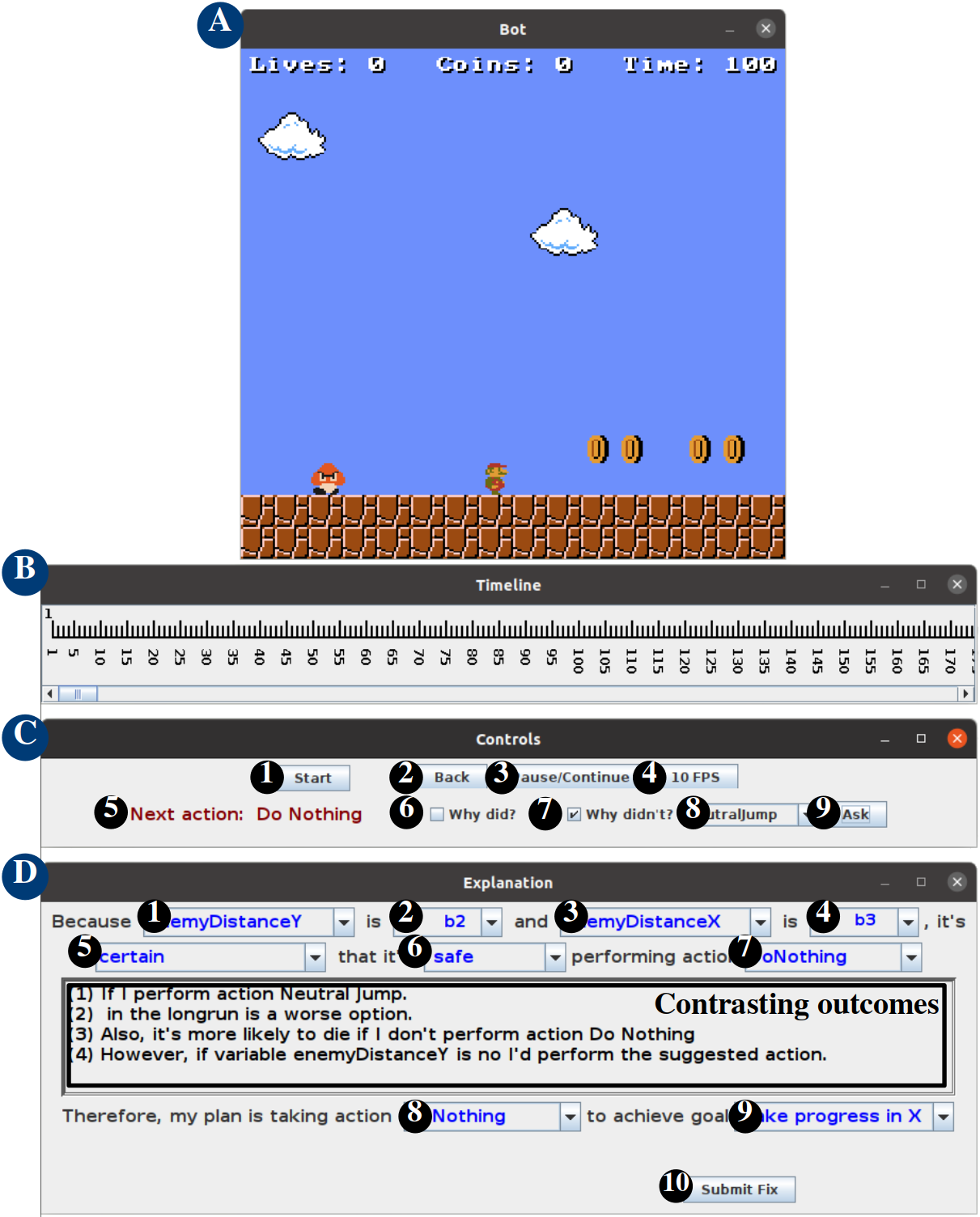}
    \caption{Our implementation of the interactive explanations framework for \textit{Super Mario Bros}.}
    \label{fig::interface}
\end{center}
\end{figure}

In Figure \ref{fig::interface}, we present the interface of our system. We introduce how to use our interface using an example of a user study in which Rey, a game enthusiast, wants to personalize the behavior of a bot that plays her favorite game -- \textit{Super Mario Bros}.

First, Rey pushes the \textbf{(1)} ``Start'' button in panel \textbf{(C)} so the game screen \textbf{(A)} appears, and a precomputed bot begins playing. Then, Rey notices that the bot always runs at an enemy and dies. She doesn't like it, so she goes back to the frame in which she considers that the bot should try to kill the enemy, and she does that by pausing the game with the button \textbf{(3)} ``Pause/Continue'' in panel \textbf{(C)} and then selecting the said frame using the timeline in panel \textbf{(B)}. 

Now Rey wants to know why the bot doesn't try to kill the enemy by jumping to the right. To do that, in panel \textbf{(C)}, she selects the \textbf{(7)} ``Why didn't'' checkbox, selects action ``jump right'' from the \textbf{(8)} ``Actions'' dropdown menu, and finally presses the button \textbf{(9)} ``Ask''. After a few seconds, Rey can read the generated explanation in panel \textbf{(D)}. 

The explanation in \textbf{(D)} gives Rey a better idea of the bot's model, so she can provide the appropriate feedback to fix its behavior. Accordingly, Rey selects the suitable features and their values using the dropdown menus from \textbf{(1)} to \textbf{(7)}. Furthermore, using the dropdown menu \textbf{(8)}, she proposes the best action the bot can take at that particular state to achieve the goal ``kill an enemy'', which she selects from the dropdown menu \textbf{(9)}. Finally, she presses button \textbf{(10)} ``Submit Fix'' in panel \textbf{(D)} and waits a few seconds to see the updated behavior of the bot by pressing the button \textbf{(3)} ``Pause/Continue'' in panel \textbf{(C)}.

Another way Rey could've asked about the bot's decision is by choosing from panel \textbf{(C)} the \textbf{(6)} ``Why did?'' checkbox and then pressing the button \textbf{(9)} ``Ask''. In this manner, the contrasting part of the explanation, shown in the zone marked as ``Contrasting outcomes'' in panel \textbf{(D)}, would compare the outcomes of performing the action in $\pi(s)$ against the second-best action the bot can take.

\section{Implementation}
Our interactive explanation framework, presented in Figure \ref{fig::interactiveExplanation}, consists of two main modules. First, we need a module that generates explanations that describe the reasons that cause the agent to select a particular action in a given state. Second, we design a module that takes the data from the interactive explanations as input to compute patches that fix the policy of the bot.

To generate both modules we need to model the problem at hand as a Markov Decision Process (MDP). In particular, an MDP defined by the tuple $\langle S,A,T,R\rangle$, where $S$ is a set of states and $A$ is a set of actions; $T \colon S \times A \times S \rightarrow [0, 1]$ is the transition function that assigns the probability of reaching state $s'$ when executing action $a$ in state $s$, and $R \colon S \times A \rightarrow \mathbb{R}$ is the reward function. Also, we assume that the agent state is outlined by a finite set of features $Z^i_t = z^i$, $i = 1, ..., N$, each taking values in a feature space $Z^i$. Furthermore, the computed policy $\pi \colon S \rightarrow A$ and value function $V \colon S \rightarrow \mathbb{R}$ are deterministic.

We assume that an expert designs an effective reward function in a way that captures the desirables states (or goals) with values $\geq 0$, and undesirable states with values $< 0$.

\subsection{Generating Explanations}
For the explanations, we use the MDP data to create explanations that characterize \textbf{(E1)} the most relevant variables in the current state to make a decision, \textbf{(E2)} the environment's dynamics, \textbf{(E3)} the short-term goal that the agent is trying to achieve, and \textbf{(E4)} contrasting outcomes between different actions.

The idea to estimate the explanation element \textbf{(E1)} is to find the feature-sets of similar states that frequently appear conditioned on a particular action $a$ determined by the policy $\pi$. That is, for a given state $s$ for which users are asking for an explanation, we compute the appearance frequency of features in similar states to $s$. We chose to use a value-based metric since it's simple and can be approximated to reduce its computational cost \cite{Lan2021}. Our similarity metric groups states that have a similar value $v$ to the reference $v_{reference}$ within the range $1.0 \pm 0.05 \times v_{reference}$. We empirically found that this range was effective for our testbed. Then, we present to users the two features with the highest number of appearances in the set of similar states.

To present users the \textbf{(E2)} environment's dynamics, we take the values of the transition function $T$ that lead to negative states $s'$ and translate them into words to express probability according to the ranges: $0.9 < T(s,a) \leq 1.0$ is \textit{certain}, $0.75 < T(s,a) \leq 0.9$ is \textit{almost certain}, $0.55 < T(s,a) \leq 0.75$ is \textit{probable}, $0.45 < T(s,a) \leq 0.55$ is \textit{changes are even}, $0.25 < T(s,a) \leq 0.45$ is \textit{probably not}, $0.10 < T(s,a) \leq 0.25$ is \textit{almost certainly not}, and $0.0 < T(s,a) \leq 0.10$ is \textit{impossible}. This information gives a sense on the lowest time-scale of environmental dynamics. 

As part of the explanation element \textbf{(E2)}, we provide information about the perceived safety by the bot. We compute this by averaging the probability of transitioning to a negative state given the current policy. Negatives states are defined in the reward function with a scalar with a value $<0$. Therefore, we describe the transition as ``Dangerous'' if it is more likely to reach a negative state or as ``Safe'' if otherwise. The agent learns by exploration the transition probabilities to safe and dangerous states.  

There's evidence that suggests that the human brain uses a hierarchy of temporal scales to represent the dynamics of the environment \cite{Kiebel2008}. The decision that the bot takes at every time-step represents the lowest level in this hierarchy. For the \textbf{(E3)} the next subgoal that the bot is pursuing, we want to give users information about the next time-scale level in the hierarchy that encodes changes in the environment every few seconds. In particular, we simulate the environment for $2$ seconds to measure the accumulated reward that the bot receives for each reward component. That is, we keep track of the individual contributions of each reward component in the reward function. We identify as the next subgoal the reward component that accumulated more value.

Finally, we also provide an interrogative debugging mechanism that allows users to ask ``why'' and ``why not'' questions, which outputs a \textbf{(E4)} comparison between the outcomes of performing different actions at a $2$ seconds time-scale. When users ask a ``why'' question, our system compares the selected action by the policy to the second-best option the bot has. To find the second-best option, we use the same simulation mechanism we implemented for \textbf{(E3)}, and from there, we choose the action that gives more reward in the near future as the second-best option. On the other hand, to use a ``why not'' question, users have to specify an action to compare to; that is, our system compares between the selected action by the policy and the proposed action by the user. For both types of question, we frame the differences between the results of \textbf{(E1)}, \textbf{(E2)}, and \textbf{(E3)} in a way that is readable for users. Furthermore, for ``why not'' questions, we search in the feature space to find which specific value would make the bot take the suggested action by the user. In Table \ref{table::examples}, we show examples of the explanations generated by our system. 

\subsection{Fixing Behaviors}
\begin{algorithm}[!t]
\caption{Computing Behavior Patch}
\label{al::1}

\KwIn{$\langle F_{fix}, g_{fix}, a_{fix} \rangle$}
\KwOut{computed $\langle \pi_{fix}, V_{fix} \rangle$}

$g_p \gets 0$\;
$g_n \gets 0$\;
$Restart_p \gets \text{Number of positive restarts}$\;
$Restart_n \gets \text{Number of failed restarts}$\;

\While{Exploration is running}
{
    $Step_i \gets \text{CurrentTimeStep()}$\;
    $s \gets \text{ObserveState()}$\;
    $a \gets$ using Eq. \ref{eq:E1}\;
    ExecuteAction($a$)\; 
    $s' \gets \text{ObserveState()}$\;
    Update $T_{fix}(s, a, s')$

    \If{Bot achieved $g_{fix}$}
    {
        $g_p \gets g_p + 1$\;
    }
    \ElseIf{Bot failed}
    {
        $g_n \gets g_n + 1$\;
        UpdateParameters(Eq. \ref{eq:E1})\;
    }

    \If{$g_p > Restart_p $ or $ g_n > Restart_n$}
    {
        StopExploration()\;
    }
}

$\langle \pi_{fix}, V_{fix} \rangle \gets$ Solve $T_{fix}$\;
\Return{$\langle \pi_{fix}, V_{fix} \rangle$}\;
\end{algorithm}

We designed our strategy for creating policy patches based on what we call an attention-based exploration process; by taking as input the information from the interactive explanations, we can drive the attention of the bot to train in a limited space of the environment to achieve a particular goal in a specific way.

For limiting the size of the training environment, we use the variables $F_{fix}$ that the user considers to be the most important for deciding a given state. Concretely, we create a training environment that fits the specification of the state for which users want to create a patch and the specified features in it \textbf{(E1)}.   

We designed Algorithm \ref{al::1} to bias the exploration process using the action $a_{fix}$ and goal $g_{fix}$ that the user suggests to the bot. The main idea of this algorithm is to restrict the time that the bot spends exploring the training environment and make it more likely to choose $a_{fix}$ over the rest of the actions in $A$. The action selection method is shown in the following:

\begin{equation}
a =
\begin{cases}
    a_{fix} 	& \textrm{if} \quad Step_i \le Bias_{steps} \\
    Random() 	& \textrm{w/prob.} \quad \psi \\
    \pi(s) 	& \textrm{w/prob.} \quad (1 - \psi) 
\end{cases}
\label{eq:E1}
\end{equation}

That is, for the first time steps $Bias_{steps}$ of the exploration process the bot will select action $a_{fix}$. After this, the bot will follow the policy $\pi(s)$ with a probability of $(1- \psi)$ or a random action from $A$ with a probability $\psi$, where $\psi$ starts with a value of $0.2$ and increases by $0.05$ every time the bot fails to achieve goal $g_{fix}$. Furthermore, the goal $g_{fix}$ becomes the only reward signal in the environment. We combine the resampling of the transition functions (using the biased exploration process) with a goal-based reward shaping mechanism.

The exploration process finishes when the bot achieves the goal $g_{fix}$ a given number of times ($3$ times in our testbed). Then, with the experience $T_{fix}$ that the bot acquires in the training environment, we compute a new policy $\pi_{fix}$ and value function $V_{fix}$ with the model-based RL algorithm in \cite{Arzate2020MarioMix}. 

We then use the policy $\pi_{fix}$ and value function $V_{fix}$ to patch the policy $\pi$ in the states defined by the features $F_{fix}$. We filter a set of states in the global policy to be updated by using the most relevant variables and their corresponding values in \textbf{E1}. In this manner, we create the set $S_{relevant}$ that includes all the states in the environment that are also defined by the current values in \textbf{E1}. Then, we integrate the policy patches into the global policy $\pi$ by updating its values in $S_{relevant}$ with those in $\pi_{fix}$. Similarly, for the global value function we apply the update function $V(s) = V(s) + 0.1 \times V_{fix}(s)$ to the states in $S_{relevant}$. 
 
Generally speaking, our policy patches aim to learn effective policies in cases of misspecified rewards, or unexact transition models, in problems that we can decompose into a sequence of subgoals.  

\section{Testbed}
We use the Mario AI Framework \footnote{https://github.com/amidos2006/Mario-AI-Framework} as a testbed, this framework is a clone of the game named \textit{Super Mario Bros}. In particular, we use the work by \cite{Arzate2020MarioMix} as a basis to implement our interactive explanation system. Therefore, our \textit{Super Mario Bros.} bot uses a model-based reinforcement learning algorithm.

To perform our experiments, we use a laptop computer with an 8th generation Intel Core i7 CPU and $16$ GB of RAM. The time needed to compute a policy patch varies depending on the situation and goes from $\sim 5$ to $\sim 30$ seconds. The computation of the explanations takes $\sim 5$ seconds for the ``Why didn't'' questions, and $\sim 15$ seconds for the ``Why did?'' questions.

\subsection{Bot Definition}
\begin{figure}[!ht]%%%Fig
\begin{center}
\includegraphics[width = 0.9 \columnwidth ]{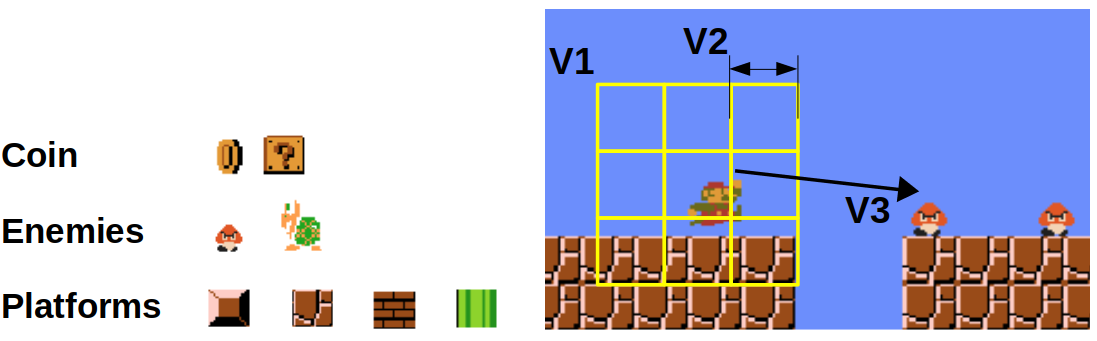}
\caption{The state representation of our bot in \textit{Super Mario Bros}.}
\label{fig::genome}
\end{center}
\end{figure}

We use the variables shown in Figure \ref{fig::genome} to represent the \textit{Super Mario Bros.} game as a Markov decision process (MDP). We use a $3 \times 3$ grid of variables \textbf{V1} that code terrain information. This grid can recognize between platforms, empty space, and coins. In particular, we use the name variable \textit{box\textbf{I}Type} for these $9$ squares that represent the value of each square in the grid. The bold \textbf{I} in the name is the index of each square. The index starts at $1$ with the top left corner, then continues to the right, and then on the next row.

Our MDP also accounts for the position (in $X$ with name variable \textit{enemyDistanceX} and $Y$ axes with name variable \textit{enemyDistanceY}) of the closest enemy (\textbf{V3}) to Mario. The position variables are discretized into $7$ values (\textit{b3, b2, b1, f1, f2, f3, no}). The values that start with a ``b'' represents when an enemy is behind Mario; while the variables that start with an ``f'' represent the opposite. The numbers (from $1$ to $3$) represent how far the enemy is where $1$ means it's very close and $3$ means it's far.

We also added a binary variable \textbf{V2} that detects whether there is a cliff close to Mario.

Additionally, we have a few variables that encode relevant information about the bot. We have a binary variable that represents if Mario can jump or not (named \textit{canJump}). Another binary variable to know if Mario is on the ground (a solid platform) or not (named \textit{onGround}). Finally, we add three binary variables that encode important states. One represents if Mario has made progress (it has got close to the goal) in the $X$ axis (the variable name is \textit{anyXProgress}). Similarly, we have another variable (\textit{anyYProgress}) that encodes if Mario has made some progress in the $Y$ axis. Finally, we have a variable to know if Mario is dead or not (named as \textit{isDead}). In Table \ref{tab::variables}, we present the names of each variable and the values they can take.

\begin{table}[htb]
\begin{center}
\begin{tabular}{cc}
\toprule
\textbf{Variable}  & \textbf{Values}\\
\midrule
box\textbf{I}Type & platform, coin, air\\
\midrule
canJump & yes, no\\
\midrule
onGround & yes, no\\
\midrule
isDead & yes, no\\
\midrule
isCliffNear & yes, no\\
\midrule
anyXProgress & yes, no\\
\midrule
anYProgress & yes, no\\
\midrule
enemyDistanceX & b3, b2, b1, f1, f2, f3, no\\
\midrule
enemyDistanceY & b3, b2, b1, f1, f2, f3, no\\
\bottomrule 
\end{tabular}
\end{center}
\caption{Variables names and their possible values in our bot definition.}
\label{tab::variables}
\end{table}

Our bot can perform $10$ different actions in total. Mario can do the next actions to the left or right of the screen: walk, run, jump, and quick jump. Furthermore, Mario can do nothing, as well as perform a neutral jump.

\section{User Study}
For our user study, we asked $13$ non-experts in reinforcement learning to fix and personalize the behavior of our bot using our proposed method. The subjects are students and staff members with an age range from $20$ to $34$, and backgrounds in design ($23.07\%$), humanities ($15.38\%$), and computer science ($61.53\%$). 

\subsection{Task}
First, we explain the capabilities of our system to users with an example. Then, we proceed to explain the task they have to complete. This task consists of fixing the policy of a bot that plays \textit{Super Mario Bros}. that stops it from finishing a given game level. In particular, we designed the game level for the test to make the bot fail at three different points; subjects had to solve all of them. Also, we encouraged the users to make at least one change in the policy to personalize the play style of the bot according to their preferences.

\subsection{The Bugs}
We created the base policy of our system by letting Mario explore the original first level of \textit{Super Mario Bros.}. Then, we designed the level that we use for our testbed with previously unseen states. Some of these novel states caused unwanted behaviors in Mario. 

The first of these bugs (\textbf{B1}) is shown in the top row of Table \ref{table::examples}. Bug \textbf{B1} makes Mario run into enemies when there are coins above him. In the row below \textbf{B1} in Table \ref{table::examples}, we show the second bug \textbf{B2}. This bug makes Mario die because when it tries to stomp on enemies, its jump trajectory is modified when it hits the platform on top of him. In the row below \textbf{B2} in Table \ref{table::examples}, we show the third bug \textbf{B3} which makes Mario infinitely jump in circles when it faces the traped enemy between the pipe and the platform. In the last row, we present how a user personalizes the bot's behavior to make it kill enemies that are behind it.

\subsection{Survey}
\begin{table*}[ht!]
    \scriptsize 
    \begin{center}
    \begin{tabular}{l m{0.27\linewidth} m{0.27\linewidth}  m{0.27\linewidth}}
    \toprule
    \textbf{Original Policy (B1)} & \textbf{Interactive Explanation} &\textbf{Contrasting Outcome Why?} & \textbf{Contrasting Outcome Why didn’t JumpRight?}\\
    \begin{minipage}{.10\textwidth}
        \includegraphics[width = 1.0 \textwidth]{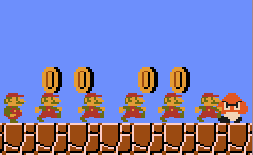}
    \end{minipage}
     & Because \hl{Box6Type} is \hl{air} and \hl{EnemyDistanceX} is \hl{f3}, it is \hl{certain} that it's \hl{safe} performing action \hl{RunRight}. Therefore, my plan is taking action \hl{RunRight} to achieve goal \hl{Make Progress in X}.
     & The second best option is doing FastJumpRight and performing it would give similar results. & If I perform action JumpRight I won't make progress in X, and in the long-run is a worse option. However, if variable box6Type is pipe I'd perform the suggested action.\\
    
     \textbf{Updated Policy} & \textbf{Fix} &\textbf{Contrasting Outcome Why?} & \textbf{Contrasting Outcome Why didn’t RunRight?}\\
    
     \begin{minipage}{.10\textwidth}
        \includegraphics[width = 1.0 \textwidth]{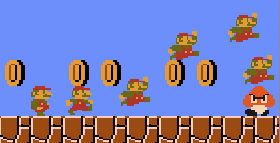}
    \end{minipage}
     & Because \hl{EnemyDistanceY} is \hl{b2} and \hl{EnemyDistanceX} is \hl{f3}, it is \hl{certain} that it's \hl{safe} performing action \hl{JumpRight}. Therefore, my plan is taking action \hl{JumpRight} to achieve goal \hl{Kill an Enemy}.
     
     & The second best option is doing Down, but I wouldn't  kill the enemy,  I wouldn't  make progress in X. And in the long-run is a worse option. Also, it's more likely to die if I don't perform action JumpRight. & If I perform action Run Right in the long-run is a worse option. Also, it's more likely to die if I don't perform action JumpRight. However, if variable EnemyDistanceY is no I'd perform the suggested action.\\
    \midrule

    \textbf{Original Policy (B2)} & \textbf{Interactive Explanation} &\textbf{Contrasting Outcome Why?} & \textbf{Contrasting Outcome Why didn’t JumpLeft?}\\
    \begin{minipage}{.10\textwidth}
        \includegraphics[width = 1.0 \textwidth]{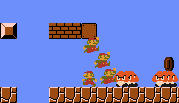}
    \end{minipage}
     & Because \hl{EnemyDistanceY} is \hl{b2} and \hl{EnemyDistanceX} is \hl{f2}, it is \hl{certain} that it's \hl{safe} performing action \hl{DoNothing}. Therefore, my plan is taking action \hl{DoNothing} to achieve goal \hl{Kill an enemy}.
     & The second best option is doing RunRight and performing it would give similar results. Also, it's more likely to die if I don't perform action DoNothing. & If I perform  action JumpLeft in the long-run is a worse option Also, it's more likely to die if I don't perform action DoNothing.\\
    
     \textbf{Updated Policy} & \textbf{Fix} &\textbf{Contrasting Outcome Why?} & \textbf{Contrasting Outcome Why didn’t JumpRight?}\\
    
     \begin{minipage}{.10\textwidth}
        \includegraphics[width = 1.0 \textwidth]{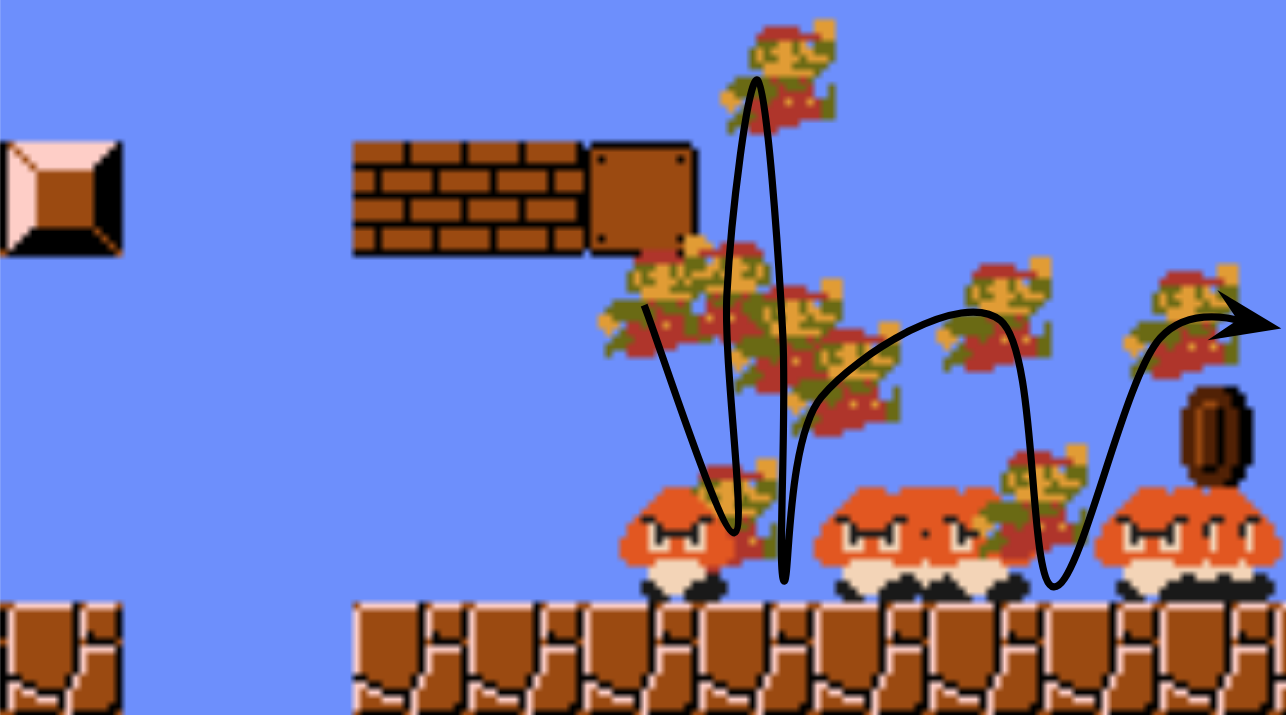}
    \end{minipage}
     & Because \hl{EnemyDistanceY} is \hl{b2} and \hl{box2Type} is \hl{ground}, it is \hl{certain} that it's \hl{safe} performing action \hl{DoNothing}. Therefore, my plan is taking action \hl{JumpLeft} to achieve goal \hl{Kill an enemy}.
     
     & The second best option is doing NeutralJump and  performing it would give similar results. Alsoo, it's more likely to die if I don't perform action DoNothing. & If I perform JumpRight I will die. However, if variable EnemyDistanceY is no I'd perform the suggestted action.\\
    \midrule

    \textbf{Original Policy (B3)} & \textbf{Interactive Explanation} &\textbf{Contrasting Outcome Why?} & \textbf{Contrasting Outcome Why didn’t FastJumpRight?}\\
    \begin{minipage}{.10\textwidth}
        \includegraphics[width = 1.0 \textwidth]{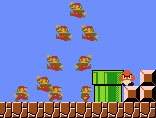}
    \end{minipage}
     & Because \hl{EnemyDistanceY} is \hl{b2} and \hl{EnemyDistanceX} is \hl{f3}, it is \hl{certain} that it's \hl{safe} performing action \hl{RunLeft}. Therefore, my plan is taking action \hl{RunLeft} to achieve goal \hl{Kill an enemy}.
     & The second best option is doing FastJumpRight and performing it would give similar results. Also, it's more likely to die if I don't perform action RunLeft. & If I perform action FastJumpRight in the long-run is a worse option. Also, it's more likely to die if I don't perform action RunLeft. However, if variable EnemyDistanceY is no I'd perform the suggestted action.\\
    
     \textbf{Updated Policy} & \textbf{Fix} &\textbf{Contrasting Outcome Why?} & \textbf{Contrasting Outcome Why didn’t JumpLeft?}\\
    
     \begin{minipage}{.10\textwidth}
        \includegraphics[width = 1.0 \textwidth]{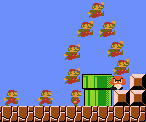}
    \end{minipage}
     & Because \hl{EnemyDistanceY} is \hl{b2} and \hl{EnemyDistanceX} is \hl{f3}, it is \hl{certain} that it's \hl{safe} performing action \hl{RunLeft}. Therefore, my plan is taking action \hl{FastJumpRight} to achieve goal \hl{MakeProgressInX}.
     
     & The second best option is doing FastJumpRight but I would't make progress in X. Also, it's more likely to die if I don't perform action JumpRight. & If I perform action JumpLeft in the long-run is a worse option. Also, it's more likely to die if I don't perform action JumpRight.\\
    \midrule

    \textbf{Original Policy} & \textbf{Interactive Explanation} &\textbf{Contrasting Outcome Why?} & \textbf{Contrasting Outcome Why didn’t NeutralJump?}\\
    \begin{minipage}{.10\textwidth}
        \includegraphics[width = 1.0 \textwidth]{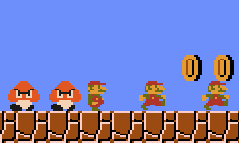}
    \end{minipage}
     & Because \hl{EnemyDistanceX} is \hl{b3} and \hl{Box5Type} is \hl{air}, it is \hl{certain} that it's \hl{safe} performing action \hl{RunRight}. Therefore, my plan is taking action \hl{RunRight} to achieve goal \hl{Make Progress in X}.
     
     & The second best option is doing FastJumpRight and performing it would give similar results. & If I perform action NeutralJump in the long-run is a worse option. Also, it's more likely to die if I don't perform action RunRight\\    

     \textbf{Updated Policy} & \textbf{Fix} &\textbf{Contrasting Outcome Why?} & \textbf{Contrasting Outcome Why didn’t RunRight?}\\
     \begin{minipage}{.10\textwidth}
         \includegraphics[width = 1.0 \textwidth]{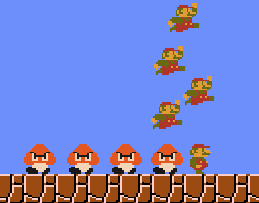}
     \end{minipage}
      & Because \hl{EnemyDistanceY} is \hl{b2} and \hl{EnemyDistanceX} is \hl{b3}, it is \hl{certain} that it's \hl{safe} performing action \hl{NeutralJump}. Therefore, my plan is taking action \hl{NeutralJump} to achieve goal \hl{Kill an Enemy}.
      
      & The second best option is doing Run Left, and  I wouldn't  make progress in X. Also, it's more likely to die if I don't perform action NeutralJump. & If I perform action RunRight in the long-run is a worse option. Also, it's more likely to die if I don't perform action NeutralJump. However, if variable EnemyDistanceY is no I'd perform the suggested action.\\    

     \bottomrule
    \end{tabular}
    \end{center}
    \caption{Explanations and fixes examples.}
    \label{table::examples}
    \end{table*}

Once the users completed the assigned task, they took a survey with the following closed-ended questions:

\begin{itemize}
\item \textbf{Q1} -- Have you ever played the game called \textit{Super Mario Bros.}? With options: Yes, No, I’ve only watched other people playing it

\item \textbf{Q2} -- Were you able to fix all the problems in the policy that you wanted? With options: Yes, No

\item \textbf{Q3} -- How close are the bot behaviors you created to what you had envisioned? With options not similar at all, some resemblance, very similar, perfect match

\item \textbf{Q4} -- How clear were the bot explanation? With options: not clear at all, a little clear, clear, perfectly clear 

\item \textbf{Q5} -- How effective were the explanations for helping you diagnosing and repairing the bot behavior? With options: not effective at all, a little effective, effective, very effective 

\item \textbf{Q6} -- Was your bot able to complete the level? With options: yes, and no
\end{itemize}

Then, we asked users the next open-ended questions:

\begin{itemize}
\item \textbf{Q7} -- Which are the parts in the interactive explanations that were more useful for you?

\item \textbf{Q8} -- How would you describe the most effective workflow using our system?

\item \textbf{Q9} -- Which parts of the contrasting outcomes explanation were most useful?
\end{itemize}

\section{Results}
In this section, we present some examples of the produced explanations and the results of the conducted survey. We invite our readers to use our system by downloading our software from this repository \footnote{https://arzate-christian.github.io/InteractiveExplanations/index.html}.

\subsection{Explanations}
In Table \ref{table::examples} we show a few instances of the explanations that our system produces. 

\subsection{Survey}
\begin{figure}[!tb]
\begin{center}
\includegraphics[width = 0.65 \columnwidth ]{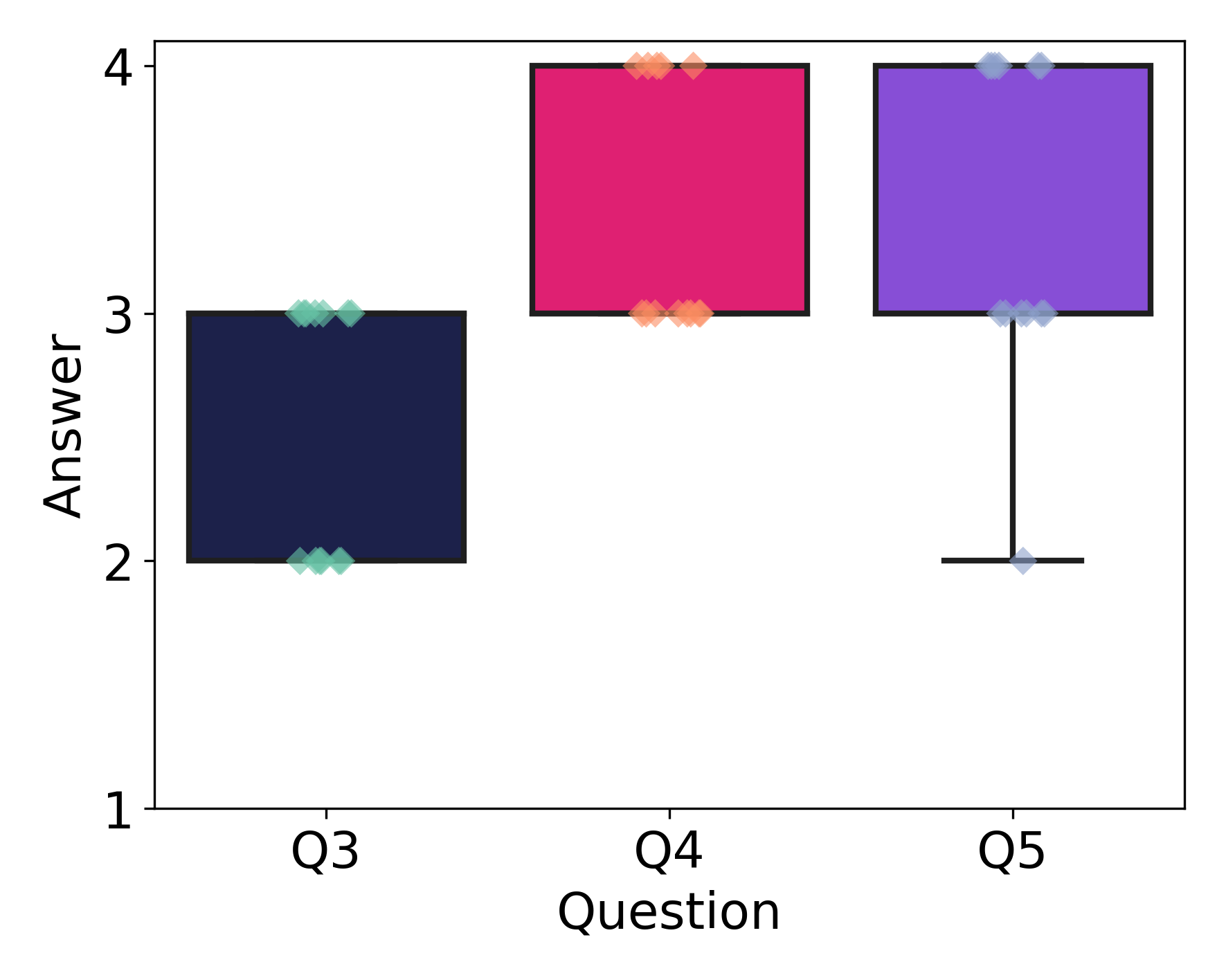}
\caption{Box plots of the survey results.}
\label{fig::boxPlot}
\end{center}
\end{figure}

Regarding \textbf{Q1}, all users stated that they have previously played \textit{Super Mario Bros.} For \textbf{Q2}, $61.53\%$ of the users were able to make all the changes in the policy they wanted, while the rest of them couldn't. For the results of questions \textbf{Q3-Q5}, we present three box plots in Figure \ref{fig::boxPlot} with the answers on a corresponding 4-level Likert scale. For the last closed-ended question (\textbf{Q6}), $92.30\%$ of the users were able to fix the policy so the bot could complete the given game level.  

The answers for the first open-ended question (\textbf{Q7}) reveal that most users ($57.14\%$) prefer using the parts where the bot describes the best action and goal (marked from $6$ to $9$ in panel \textbf{(D)} in Figure \ref{fig::interface}); while $42.85$ of the users mentioned that they prefer using the variables that describe the environment (marked from $1$ to $4$ in panel \textbf{(D)} in Figure \ref{fig::interface}) and the parts where the bot describes the best action and goal. Only one of the users mentioned the contrasting outcomes of the ``Why did?'' question as the most useful to diagnose and repair the bot's behavior. 

For question \textbf{Q8}, we can summarize the described workflow of the users as follows: (1) pause at a frame close to the behavior which needs some fix, (2) scroll to the closest frame possible in the timeline, (3) ask the bot for an explanation, (4) read the explanation, (5) modify the dropdown menus, (6) submit a fix, (7) test and iterate. It's worth mentioning that most users mentioned that it's very important to find the exact frame where the bot starts to perform an unwanted action. Besides, $69.23\%$ of the users preferred to use only ``Why didn’t?'' questions so they could compare the outcomes with the action they believe was the best to perform.

Finally, for question \textbf{Q9}, the majority of users ($38.88\%$) found most useful the information about the changes needed in the environment so the bot would choose to perform the suggested action in the ``Why didn’t?'' question. The parts related to the information about the goal that the bot wouldn't achieve if it performs the suggested action, the information regarding the danger, and the rewards in the long-run have the same preference percentage at $16.66\%$. The second-best option in the ``Why did?'' has an $11.11\%$ of popularity among users.

\section{Discussion}
The results of our user test are promising; users felt comfortable using our interactive explanations system. Also, they found it natural to use and effective at fixing the bot's behavior. Furthermore, users created novel playstyles and fixed multiple bugs besides those we asked to fix for the user test. For instance, users taught the bot how to kill enemies when its jump trajectory is limited by a platform above it.

Users spent using our platform between $30$ to $190$ minutes. The mean time spent fixing the $3$ bugs of the task was $24.67$ minutes. Only one user couldn't fix a bug (\textbf{B3}) that stopped Mario from finishing the level.

To shed some light on the cause of the bugs, we tried to fix them using traditional reward shaping. We created $8$ new bots, and our system took between $3$ and $7$ hours to find a policy for them ($\tilde{x} = 6.175$). From all the new bots, $2$ of them couldn't solve any bug, and the rest of them solved all bugs except the bug \textbf{B1}. We believe that an unexact transition model caused bug \textbf{B1}, while a misspecified rewards function caused the rest. This evidence suggests that our method can fix bugs caused by both misspecified rewards and unexact transition models. However, we require to implement a mechanism to better understand the cause of bugs and how our algorithm solves them.     

\section{Limitations}
One limitation of our method is that we require a base policy for which we can create patches that make adjustments to the base behavior but this can make it difficult to make a global change. Besides, if users create patches with contradictory goals that affect similar states, these changes can create unwanted behaviors in the bot. We can mitigate the latter by giving the users the option of specifying whether a patch should be globally applied or only for the given place.

Some users were not able to make all the changes to the policy that they wanted. They were limited by the representation of the environment of the bot and the time-scale of our model. To solve this, we would need to give users the ability to create arbitrary (sub)goals and create transition models at a higher level time scales.

Another disadvantage is that we need hand-engineered MDPs to generate the explanations which is time-consuming. One way to reduce the time of experts is using an inverse RL algorithm \cite{Ng1999} to find a base policy and reward function for the problem at hand. To facilitate the generation of explanations with our framework, we could implement object grounding techniques \cite{Roesler2019} so non-experts in RL could teach the bot the meaning of objects and actions using natural language.

\section{Conclusions}
In this paper, we introduced a novel interaction mechanism for diagnosing and repairing agent behaviors through editable explanations in natural language templates. The main advantage of our method is that it enables a two-way communication channel between users and bots. Furthermore, in our user test, we found out that our editable explanations framework is effective at providing clear explanations that facilitate users to patch the behavior of the bot with a fast interaction cycle.

\section*{Acknowledgements}
This work was supported by JST CREST Grant Number JP-MJCR17A1, Japan. Additionally, we would like to thank the reviewers of our paper. Their kind suggestions helped to improve and clarify this manuscript.

\bibliographystyle{IEEEtran}
\bibliography{Bibliography}

\end{document}